\documentclass[10pt]{article}
\usepackage[cp866]{inputenc}
\begin{document}

\centerline {{\Large\bf Conservation laws.}}
\centerline {{\Large\bf Generation of physical fields.}}
\centerline {{\Large\bf Principles of field theories}}

\centerline {\bf L.I. Petrova}
\centerline{{\it Moscow State University, Russia, e-mail: ptr@cs.msu.su}}
\renewcommand{\abstractname}{Abstract}
\begin{abstract}

In the paper the role of conservation laws in evolutionary processes,
which proceed in material systems (in material media) and lead to
generation of physical fields, is shown using skew-symmetric 
differential forms.

In present paper the skew-symmetric differential forms on
deforming (nondifferentiable) manifolds were used in addition to
exterior forms, which have differentiable manifolds as a
basis. Such skew-symmetric forms [1], whose existence was established
by the author (and which were named evolutionary ones since they
possess evolutionary properties), as well as the closed exterior
forms, describe the conservation laws. But in contrast to exterior
forms, which describe conservation laws for physical fields, the
evolutionary forms correspond to conservation laws for material 
systems.

The evolutionary forms possess an unique peculiarity, namely, the
closed exterior forms are obtained from these forms. It is just
this that enables one to describe the process of generation of
physical fields, to disclose connection between physical fields and
material systems and to resolve many problems of existing field
theories.
\end{abstract}

{\large\bf Introduction}

Skew-symmetric differential forms possess a peculiarity (which
does not possesses any other mathematical apparatus), namely, they
describe conservation laws. It is known that closed exterior
differential forms are conservative quantities (the differential
of closed form vanishes). And closed inexact exterior forms are
conservative object: quantities conserved on structures (more precisely, on
pseudostructures, which are described by dual forms). Physical 
structures, from which physical fields are formed and to which 
the conservation law is assigned, are such
conservative objects. From this one can see that
closed exterior inexact forms describe conservation laws for
physical fields. These are conservation laws for physical fields that 
can be named exact ones. In field theories by a conception 
of "conservation laws" is meant just such conservation laws.

However, in physics and mechanics of continuous media the
conception of "conservation laws" is related to conservation laws,
which can be called balance ones. These are conservation laws for
material systems (continuous media) - conservation laws for
energy, linear momentum, angular momentum, and mass, which
establish a balance between physical quantities and external
actions to the system. They are described by differential
equations. And in this case from differential equations it follows
a relation, which includes a skew-symmetric differential form.
This skew-symmetric form [1] possesses the evolutionary properties and from 
that closed exterior forms corresponding to exact conservation laws are 
obtained. The passing from the evolutionary form, which correspond to 
balance conservation laws for material systems, to closed exterior forms, 
which correspond to conservation laws for physical fields, describes the 
process of generating physical fields by material systems.

The connection between physical fields and material systems underlines 
the fact that fields theories describing physical fields have to be 
based on the principles that specify material system.

\subsection*{1. Conservation laws}

It has been noted that in mathematical physics there are two types
of conservation laws, namely, conservation laws, which can be
called exact ones, and balance conservation laws.

The exact conservation laws are related to physical fields.

The balance conservation laws are conservation laws for material
systems (material media). 
{\footnotesize [Material system is a variety of elements that possess 
internal structure and interact to one another. Thermodynamic and gas
dynamical systems, systems of charged particles, cosmic systems, systems
of elementary particles and others are examples of material systems.
Examples of elements that constitute material system
are electrons, protons, neutrons, atoms, fluid particles, cosmic objects
and others.]}

Below it will be shown that there exists a connection between balance 
and exact conservation laws which points to a connection between 
material systems and physical fields.

\bigskip

{\large\bf Exact conservation laws}.

The closed exterior differential forms describes exact conservation
laws.

From the closure conditions of exterior differential
form (vanishing the form differential)
$$
d\theta^k=0\eqno(1)
$$
one can see that the closed exterior differential form is a conservative
quantity ($\theta^k$ is exterior differential form of degree $k$ - ($k$-form)).
This means that it can correspond to conservation law,
namely, to some conservative physical quantity.

If the exterior form is a closed inexact form, i.e. is closed only
on pseudostructure, the closure condition is written as
$$
d_\pi\theta^k=0\eqno(2)
$$
And the pseudostructure $\pi$ obeys the condition
$$
d_\pi{}^*\theta^k=0\eqno(3)
$$
where ${}^*\theta^k$ is a dual form.

From conditions (2) and
(3) one can see that the closed exterior form and the dual form 
constitute a conservative object, namely, a quantity that is
conservative on the pseudostructure. Hence, such an object can
correspond to some conservation law.

The closure conditions for the exterior differential form
($d_{\pi }\,\theta ^k\,=\,0$)
and the dual form ($d_{\pi }\,^*\theta ^k\,=\,0$) are
mathematical expressions of the exact conservation law.

The pseudostructure (dual form) and
the conservative quantity (closed exterior form) define a
differential-geometrical structure (which is an example of G-Structure).
It is evident that such differential-geometrical structure corresponds
to exact conservation law.

Below it will be shown that physical structures, which form physical fields,
are such differential-geometrical structures.

The mathematical expression for exact conservation law and its
connection with physical fields can be schematically written in
the following manner:
$$
\def\\{\vphantom{d_\pi}}
\cases{d_\pi \theta^k=0\cr d_\pi {}^{*\mskip-2mu}\theta^k=0\cr}\quad
\mapsto\quad
\cases{\\\theta^k\cr \\{}^{*\mskip-2mu}\theta^p\cr}\quad\hbox{---}\quad
\hbox{physical structures}\quad\mapsto\quad\hbox{physical fields}
$$

It can be shown that field theories, i.e. theories
that describe physical fields, are based on the invariant and metric
properties of closed exterior differential and dual forms
that correspond to exact conservation laws.

\bigskip
{\large\bf Balance conservation laws}.

In mechanics and physics of material systems (of continuous media)
the equations of balance conservation laws are used for
description of physical quantities, which specify the behavior of
material systems. But the balance conservation laws not only
define the variation of physical quantities. Their role is much
wider. They control evolutionary processes in material systems
that are accompanied by an origin of physical structures.

Evolutionary processes are described by the relations that are obtained
from the equations of balance conservation laws.

\bigskip

The equations of balance conservation laws are differential (or
integral) equations that describe the variation of functions corresponding
to physical quantities [2-4].
The functions for equations of material media sought are usually
functions which relate to such physical quantities like a particle
velocity (of elements), temperature or energy, pressure and
density. Since these functions relate to one material system, it
has to exist a connection between them. This connection is
described by the state-function.

From the equations of balance conservation laws one gets the relation 
for state-function containing unclosed skew-symmetric differential
form. Such a relation, which appears to be nonidentical one (since
this relation includes unclosed form), just describes evolutionary
processes in material media.

The derivation of this relation can be demonstrated by the example
of equations that describe the balance conservation laws for
energy and linear momentum.

We introduce two frames of reference: the first is an inertial one
(this frame of reference is not connected with material system), and
the second is an accompanying
one (this system is connected with the manifold built by
trajectories of material system elements). The energy equation
in inertial frame of reference can be reduced to the form:
$$
\frac{D\psi}{Dt}=A_1\eqno(4)
$$
where $D/Dt$ is the total derivative with respect to time, $\psi $ is the
functional of the state that specifies the material system, $A_1$ is the
quantity that depends on specific features of the system and on external
energy actions onto the system.
{\footnotesize \{The action functional, entropy, wave function
can be regarded as examples of the functional $\psi $. Thus, the equation
for energy presented in terms of the action functional $S$ has a similar form:
$DS/Dt\,=\,L$, where $\psi \,=\,S$, $A_1\,=\,L$ is the Lagrange function.
In mechanics of continuous media the equation for
energy of ideal gas can be presented in the form [4]: $Ds/Dt\,=\,0$, where
$s$ is entropy. \}}

In the accompanying frame of reference the total derivative with respect to
time converts into the derivative along trajectory. Equation
(4)
is now written in the form
$$
{{\partial \psi }\over {\partial \xi ^1}}\,=\,A_1 \eqno(5)
$$
here $\xi^1$ is the coordinate along trajectory.

In a similar manner, in the
accompanying frame of reference the equation for linear momentum appears
to be reduced to the equation of the form
$$
{{\partial \psi}\over {\partial \xi^{\nu }}}\,=\,A_{\nu },\quad \nu \,=\,2,\,...\eqno(6)
$$
where $\xi ^{\nu }$ are the coordinates in the direction normal to
trajectory, $A_{\nu }$ are the quantities that depend on specific
features of the system and external (with respect to local domain)
force actions.

Eqs. (5), (6) can be convoluted into the relation
$$
d\psi\,=\,A_{\mu }\,d\xi ^{\mu },\quad (\mu\,=\,1,\,\nu )\eqno(7)
$$
where $d\psi $ is the differential
expression $d\psi\,=\,(\partial \psi /\partial \xi ^{\mu })d\xi ^{\mu }$
(the summation over repeated indices is implied).

Relation (7) can be written as
$$
d\psi \,=\,\omega \eqno(8)
$$
here $\omega \,=\,A_{\mu }\,d\xi ^{\mu }$ is the skew-symmetric differential
form of first degree.

Since the equations of balance conservation laws are evolutionary
ones, the relation obtained is also an evolutionary relation.

Relation (8) was obtained from the equation of balance
conservation laws for
energy and linear momentum. In this relation the form $\omega $ is that of
first degree. If the equations of balance conservation laws for
angular momentum be added to the equations for energy and linear momentum,
this form in the evolutionary relation will be a form of second degree.
And in  combination with the equation of balance conservation law
of mass this form will be a form of degree 3.

Thus, in general case the evolutionary relation can be written as
$$
d\psi \,=\,\omega^p \eqno(9)
$$
where the form degree  $p$ takes values $p\,=\,0,1,2,3$..
(The evolutionary
relation for $p\,=\,0$ is analogue to that in differential forms, and it was
obtained from interaction of energy and time.)
In relation (8) the form $\psi$ is a form of zero degree. And in relation
(9) the form $\psi$ is a form of $(p-1)$ degree.

Let us show that {\it the evolutionary relation  obtained from the 
equation of balance conservation laws proves to be nonidentical}.

To do so we shall analyze relation (8).

In the left-hand side of evolutionary relation (8) there is the
differential that is a closed form. This form is an invariant
object. The right-hand side of relation (8) involves the
differential form $\omega$, that is not an invariant object since
in real processes, as it will be shown below, this form proves to
be unclosed.

For a form to be closed the differential of the form or its commutator
must be equal to zero.
Let us consider the commutator of the form
$\omega \,=\,A_{\mu }d\xi ^{\mu }$.
The components of commutator of such a form can be written as follows:
$$
K_{\alpha \beta }\,=\,\left ({{\partial A_{\beta }}\over {\partial \xi ^{\alpha }}}\,-\,
{{\partial A_{\alpha }}\over {\partial \xi ^{\beta }}}\right )\eqno(10)
$$
(here the term  connected with the manifold metric form
has not yet been taken into account).

The coefficients $A_{\mu }$ of the form $\omega $ have been obtained either
from the equation of balance conservation law for energy or from that for
linear momentum. This means that in the first case the coefficients depend
on energetic action and in the second case they depend on force action.
In actual processes energetic and force actions have different nature and appear
to be inconsistent. The commutator of the form $\omega $ constructed from
derivatives of such coefficients is nonzero.
This means that the differential of the form $\omega $
is nonzero as well. Thus, the form $\omega$ proves to be unclosed and is not
a measurable quantity.

This means that relation (8) involves a noninvariant term.
Such a relation cannot be an identical one.
Hence, without a knowledge of concrete expression for the form
$\omega$, one can argue that for actual processes the relation obtained
from the equations corresponding to balance conservation laws proves
to be nonidentical.

Similarly it can be shown that general relation (9) is also nonidentical. 
(The analysis of some particular equations of balance conservation
laws and relevant evolutionary relations are presented in papers [1]).

{\footnotesize \{The peculiarities of nonidentity of evolutionary relation
are connected with the differential form $\omega^p $ that enters into this
relation.
The form $\omega^p $ in evolutionary relation is a skew-symmetric differential
form. However, this form is not {\it exact} one. Unlike to exterior form,
whose basis is a differential manifold, this form is defined on deforming
(nondifferentiable) manifold. (About properties of such skew-symmetric form
one can read, for example, in paper [1]). The peculiarity of 
skew-symmetric forms defined on such manifold is the fact that their 
differential depends on the basis. The commutator of such form includes 
the term that is connected with a differentiating the basis. This can
be demonstrated by the example of a skew-symmetric form of first-degree.

Let us consider the first-degree form
$\omega=a_\alpha dx^\alpha$. The differential of this form can
be written as $d\omega=K_{\alpha\beta}dx^\alpha dx^\beta$, where
$K_{\alpha\beta}=a_{\beta;\alpha}-a_{\alpha;\beta}$ are
components of commutator of the form $\omega$, and
$a_{\beta;\alpha}$, $a_{\alpha;\beta}$ are covariant
derivatives. If we express the covariant derivatives in terms of
connectedness (if it is possible), they can be written
as $a_{\beta;\alpha}=\partial a_\beta/\partial
x^\alpha+\Gamma^\sigma_{\beta\alpha}a_\sigma$, where the first
term results from differentiating the form coefficients, and the
second term results from differentiating the basis. If we substitute
the expressions for covariant derivatives into the formula for
commutator components, we obtain the following expression
for commutator components of the form $\omega$:
$$
K_{\alpha\beta}=\left(\frac{\partial a_\beta}{\partial
x^\alpha}-\frac{\partial a_\alpha}{\partial
x^\beta}\right)+(\Gamma^\sigma_{\beta\alpha}-
\Gamma^\sigma_{\alpha\beta})a_\sigma\eqno(11)
$$
Here the expressions
$(\Gamma^\sigma_{\beta\alpha}-\Gamma^\sigma_{\alpha\beta})$
entered into the second term are just components of commutator of
the first-degree metric form that specifies the manifold
deformation and hence is nonzero. (In the commutator of exterior
form, which is defined on differentiable manifold, the second term
absents: the connectednesses are symmetric, that is, the
expression
$(\Gamma^\sigma_{\beta\alpha}-\Gamma^\sigma_{\alpha\beta})$
vanishes). [It is well-known that the metric form commutators of
first-, second- and third degrees specifies, respectively,
torsion, rotation and curvature.]

The skew-symmetric form in evolutionary relation is defined on the
manifold made up by trajectories of the material system elements.
Such a manifold is a deforming manifold. The commutator of
skew-symmetric form  defined on such manifold includes the metric
form commutator being nonzero. (In expression (10) one more term
connected with the torsion of accompanying manifold on which the
form $\omega \,=\,A_{\mu }d\xi ^{\mu }$) is defined will appear.
The commutator of such skew-symmetric form cannot be equal to
zero. And this means that evolutionary skew-symmetric form, which
enters into evolutionary relation, cannot be closed.

Nonclosure of evolutionary form and the properties of commutator of such
form define properties and peculiarities of the relation obtained from
the equations of balance conservation laws.\}}

\bigskip
Below it will be shown that the properties and peculiarities of nonidentical
evolutionary relation enables one to understand the mechanism of evolutionary
processes in material systems and the mechanism of generation of physical fields.

\subsection*{2. Connection between physical fields and material systems.
Generation of physical fields}

The nonidentity of evolutionary relation means that the balance
conservation law equations are inconsistent (nonconjugated). This
reflects the properties of the balance conservation laws
that have a governing importance for the evolutionary processes in
material media, namely, their {\it noncommutativity}.

The noncommutativity of balance conservation laws causes the fact that
the material system state appears to be nonequilibrium one.

It is evident that, if the balance conservation laws be commutative,
the evolutionary relation would be identical and from that it would be
possible to get the differential $d\psi $ and find the state-function,
and
this would indicate that the material system is in equilibrium state.

However, as it has been shown, in real processes the balance conservation laws
are noncommutative. The evolutionary relation is not identical and from
this relation one cannot get the differential $d\psi $. This means that
the system state is nonequilibrium. It is evident that
the internal force producing such nonequilibrium state is
described by the evolutionary form commutator. Everything that
gives contribution to the commutator of the form $\omega^p $
leads to emergence of internal force.

Nonidentical evolutionary relation also describes how the state of
material system changes. This turns out to be possible due to the
fact that the evolutionary nonidentical relation is a selfvarying
one. This relation includes two objects one of which appears to be
unmeasurable.  The variation of any object of the relation in some
process leads to variation of another object and, in turn, the
variation of the latter leads to variation of the former. Since
one of the objects is a unmeasurable quantity, the other cannot
be compared with the first one, and hence, the process of mutual
variation cannot stop. This process is governed by the
evolutionary form commutator, that is, by interaction between the
commutator made up by derivatives of the form itself and by metric
form commutator of deforming manifold made up by trajectories of elements
of material system.

Selfvariation of nonidentical evolutionary relation points to the
fact that the nonequilibrium state of material system turns out
to be selfvarying. The state of material system changes but holds
nonequilibrium during this process.

\bigskip
During selfvariation of evolutionary relation it can be realized
conditions when an inexact (closed {\it on pseudostructure}) 
exterior form is obtained from evolutionary form. This leads to the
fact that from nonidentical evolutionary relation it will be
obtained an identical (on pseudostructure) relation, and this
points to the transition of material system from nonequilibriun
state to locally equilibrium state.

The transition from unclosed evolutionary form to closed exterior form
is possible only as degenerate transformation,
namely, a transformation that does not conserve the differential.
The conditions of degenerate transformation are those that
determine the direction on which interior (only along a given
direction) differential of evolutionary form vanishes.
These are conditions that define the pseudostructure, i.e.
the closure conditions of dual form, and lead to realization of
the exterior form closed on pseudostructure.
{\footnotesize [The conditions of degenerate transformation are some
symmetries. Such conditions can be due to degrees of freedom of
material system (like, for example, translation, rotation, oscillation
and so on) that are realized while selfvarying of nonequilibrium state
of material system.]}

As it has been already mentioned, the differential of the evolutionary
form $\omega^p$ involved into nonidentical relation (9) is nonzero.
That is, $d\omega^p\ne 0 $.
If the conditions of degenerate transformation are realized, it will take place
the transition 

$d\omega^p\ne 0 \to $ (degenerate transformation) $\to d_\pi \omega^p=0$, 
$d_\pi{}^*\omega^p=0$

The relations obtained
$$d_\pi \omega^p=0,  d_\pi{}^*\omega^p=0 \eqno(12)$$
are closure conditions for exterior inexact form and for dual form.
This means that
it is realized an exterior form closed on pseudostructure.

In this case, on the pseudostructure $\pi$ evolutionary relation (9) converts
into the relation
$$
d_\pi\psi=\omega_\pi^p\eqno(13)
$$
which proves to be an identical relation. Since the form
$\omega_\pi^p$ is a closed one, on the pseudostructure this form
turns out to be a differential. There are differentials in the
left-hand and right-hand sides of this relation. This means that
the relation obtained is an identical one.

From identical relation one can obtain the state differential
and find the state function, and this points to the material system state
is a equilibrium state. But this state is realized only locally since
the state differential is interior one defined exclusively on pseudostructure.
({\it The total
state of material system turns out to be nonequilibrium} because
the evolutionary relation itself remains to be nonidentical one.)

Relation (13) holds the duality. The left-hand side of relation
(13) includes the differential, which specifies material system
and whose availability points to the locally-equilibrium state of
material system. And the right-hand side includes the closed inexact
form, which is a characteristics of physical fields. The closure
conditions (12) for exterior inexact form correspond to
conservation law, i.e. to a conservative on pseudostructure
quantity, and describe a differential-geometrical structure. These
are such structures (pseudostructures with conservative
quantities) that are physical structures formatting physical
fields. Massless particles, charges, structures made up by eikonal
surfaces and wave fronts, and so on are examples of physical structures.

The transition from nonidentical relation (9) obtained from balance
conservation laws to identical relation (13) means the following.
Firstly, the existence of state differential (left-hand side of
relation (13)) points to transition of material system from
nonequilibrium state to locally-equilibrium state. And, secondly,
the emergence of closed (on pseudostructure) inexact exterior form
(right-hand side of relation (13)) points to origination of physical
structure (from which physical fields are made up).

The duality of identical relation also explains the duality of
nonidentical evolutionary relation. On the one hand, the
evolutionary relation describes the evolutionary process in
material systems, and, on the other, describes the process of
emergence of physical structures and generating physical fields.

The emergence of physical structures in evolutionary process
reveals in material system as an advent of certain observable
formations, which develop spontaneously. Such formations and their
manifestations are fluctuations, turbulent pulsations, waves, vortices,
and others. It appears that structures of physical fields and the
formations of material systems observed are a manifestation of the same
phenomena. The light is an example of such a duality. The light
manifests itself in the form of a massless particle (photon) and of
a wave.

By sequential integrating of the evolutionary relation the closed inexact
exterior forms of degree $k$ are obtained from the evolutionary form of
degree $p$, where $k$ ranges
from $p$ to $0$. In this case the pseudostructures
of dimensions $(n+1-k)$ correspond to closed forms of degree $k=p$,
$k=p-1$, \dots, $k=0$.
{\footnotesize \{Under
degenerate transformation from the nonidentical evolutionary
relation one obtains a relation being identical on pseudostructure,
that can be integrated. The relation obtained after integration proves
to be nonidentical as well.
By sequential integrating the nonidentical relation of degree $p$ (in
the case of realization conditions of corresponding degenerate transformations
and forming the identical relation), one can get a closed (on the
pseudostructure) exterior forms of appropriate degrees.\}}

The parameters of evolutionary and exterior forms $p$, $k$, $n$ enables one
to introduce the classification of physical structures that defines a type of
physical structures
and, accordingly, of physical fields and interactions (See, Appendix).

Since the physical structures are generated by material media, their
characteristics are specified by characteristics of material systems,
by the characteristics of evolutionary form and of closed exterior form realized 
and by the quantity of nonvanishing commutator of evolutionary form [1].
(Specifically, the closed exterior form realized defines such a characteristics 
like a charge).

\bigskip
In conclusion of this section it should be emphasized the role of
conservation laws in generation of physical fields.

The nonidentity of evolutionary relation obtained from the equations 
that describe conservation laws for material systems (material media) 
points to a noncommutativity of these conservation laws, which are
balance ones rather then exact. The noncommutativity of
conservation laws leads to evolutionary processes in material
media, which gives rise to generation of physical fields. The
generation of physical fields is caused by the fact that due to
availability of material system degrees of freedom the conditions, 
under which the
balance conservation laws locally (only under these conditions)
commutate and become an exact conservation laws, are realized in
material system. And this points to emergence of physical structures
from which physical fields are formed.

\bigskip
The connection between physical fields and material systems has to be
taken into account in field theories as well. 

\subsection*{3. Basic principles of existing field theories}

It can be shown that the field theories are based on invariant
and metric properties of closed exterior (inexact) differential and
dual forms, which correspond to exact conservation laws.

The properties of closed exterior and dual forms, namely, invariance,
covariance, conjugacy, and duality, lie at the basis of the group,
structural and other invariant methods of field theories.

The nondegenerate transformations of field theory are
transformations of closed exterior form - nondegenerate transformations
conserving the differential.

These are gauge transformations for spinor, scalar, vector, and tensor
fields, which are transformations of closed ($0$-form),
($1$-form), ($2$-form) and ($3$-form) respectively.

The gauge, i.e. internal, symmetries of field theory
(corresponding to gauge transformations) are those of closed exterior
forms. The external symmetries of the equations of field theory are
symmetries of closed dual forms.

The field theory operators are connected with nondegenerate
transformations of exterior differential forms [5].

It can be shown that the equations of existing field theories are
those obtained on the basis of the properties of exterior form
theory.

In equations of existing field theories the closure conditions of exterior or
dual forms are laid. The postulates on which the equations of existing
field theories are such conditions. Closed inexact or dual forms are solutions
of the field-theory equations.

The Hamilton formalism is based on the properties of closed exterior
form of the first degree and corresponding dual form.
From the set of Hamilton equations and from corresponding field equation
the identical relation with exterior form of first degree, namely,
the Poincare invariant $ds\,=-\,H\,dt\,+\,p_j\,dq_j$ is obtained.

The Schr\H{o}dinger equation in quantum mechanics is an analog to
field equation, where the conjugated coordinates are replaced by
operators connected with the exterior forms of zero degree. 
The Heisenberg equation corresponds to the closure
condition of dual form of zero degree. Dirac's {\it bra-} and
{\it cket}- vectors made up a closed exterior form of zero degree.
It is evident that the relations with closed skew-symmetric
differential and dual forms of zero degree correspond to quantum
mechanics.

The properties of closed exterior form of second degree (and dual
form) lie at the basis of the electromagnetic field equations.
The strength tensor $F_{\mu\nu}$ in the Maxwell equations obeys the identical relations
$d\theta^2=0$, $d^*\theta^2=0$ [5],
where $\theta^2=\frac{1}{2}F_{\mu\nu}dx^\mu dx^\nu$ is a closed exterior
form of second degree.

Closed exterior and dual forms of third degree correspond to
gravitational field. The Einstein equation is a
relation in differential forms that relates the
differential of dual form of first degree (Einstein's tensor) and
the closed form of second degree -- the energy-momentum tensor.
And it can be shown that Einstein's equation is obtained from the relations
which connect the differential forms of third degree [6].

One can recognize that equations of field theories, as well as the gauge
transformations and symmetries, are connected with closed exterior forms of
given degree. This enables one to introduce a classification of physical
fields and interactions according to the degree of closed exterior form.
This shows that there exists a commonness between field theories
describing physical fields of different types. The degree of closed
exterior forms is a parameter that integrates fields theories into
unified field theory.

Thus, it is evident that field theories are based on the properties of
closed exterior and dual forms.

However, in existing field theories there are no answers to following questions.

1. From what one may take closed exterior forms that correspond to
conservation laws and on which properties field theories are based?

2. What defines the degree of closed exterior forms that can be a parameter
of unified field theory? Why this parameter varies from $0$ to $3$?

3. By what the quantum character of field theories is conditioned?

4. By what the symmetries and transformations of field theories are
conditioned?

The evolutionary skew-symmetric forms enable one to answer these
questions.

It was shown that the evolutionary forms allow to describe the process
of generation of physical fields, which discloses a connection between
physical fields and material systems. And this points to the fact that 
at the basis of field theories, i.e. theories that describe
physical fields, it has to lie the principles taking into account 
the connection of physical fields and material systems.

\subsection*{4. On foundations of field theory}
In the second section it had been shown that
closed exterior forms, which correspond to conservation laws for
physical fields and on which properties the theories describing physical
fields are based,
are connected with the equations for material systems. These closed exterior 
forms are obtained
from evolutionary forms in nonidentical relation derived from the equations of balance
conservation laws for material systems.

And it was shown that the degrees of relevant closed forms are connected with the
degree $p$ of evolutionary form in the nonidentical relation.
{\footnotesize (It should be recalled that the degree of evolutionary form $p$
is connected with the number of interacting balance conservation laws for material
media and can take the values $0, 1, 2, 3$. In this case from the nonidentical
relation with evolutionary form of degree $p$ the closed (inexact) forms of degrees
$k$, which can take the values $p, p-1, ..., 0$, are obtained in the
process of sequential integrating (if the degenerate transformations are realized).)}

Since physical fields, as it had been shown, are formed up by physical
structures, this means that physical fields are discrete ones rather then
continuous. The exterior closed forms corresponding to conservation laws
are {\it inexact } forms because they are obtained only under degenerate
transformations. Hence, the conservation laws corresponding to physical 
fields are satisfied on physical structures only. (For physical fields 
be continuous ones, the exact exterior forms must correspond to these fields).

The discreteness of physical fields points to the fact that field theories mast be
quantum ones.

By what symmetries and transformations of field theories are conditioned?

The external symmetries of the equations of field theory are
symmetries of closed dual forms. It had been shown that the
symmetries of dual forms are connected with the condition of
degenerate transformations, which are realized in the process of
selfvariation of material system. It is clear that such symmetries
are conditioned by degrees of freedom of material system
(translational, rotational, oscillatory and so on). Hence, the
external symmetries of the equations of field theory are also
conditioned by degrees of freedom of material system.

The gauge, i.e. internal, symmetries of the field theory
(corresponding to the gauge transformations) are those of closed
exterior forms. The symmetries of closed exterior forms are
symmetries of differentials of skew-symmetric forms (the closure
conditions of the form, namely, vanishing the form differential,
are connected with these forms). The differential of closed
inexact form obtained from evolutionary form (and corresponding to
physical structure) is an interior, being equal to zero,
differential of evolutionary form. This differential is obtained
from the evolutionary form coefficients and therefore is connected
with the characteristics of material system. As the result, the
symmetries of closed exterior forms, and, consequently, the
interior symmetries of field theory, are defined by the
characteristics of material system.

The symmetry of dual forms lead to degenerate transformations, i.e.
to going from evolutionary forms (with nonzero differential) to
closed exterior forms (with the differential being equal to zero). And the symmetries of
exterior forms lead to nondegenerate transformations, namely, to transitions
from one closed form to another closed form. Thus, it appears that degenerate and
nondegenerate transformations are interrelated.

This is also valid for transformations in field theory, since the interior
and exterior symmetries of field theories are connected with the symmetries
of closed exterior and dual forms. Thus, it turns out that the gauge nondegenerate
transformations of field theories are connected with degenerate transformations.
The transformations of field-theory equations, to which the exterior symmetries
correspond, are such degenerate transformations.

What are general properties of the equations of field theories?

The equations of fields theories, which describe physical fields,
must be connected with the equations that describe material systems,
since  material systems generate physical fields.

The equations of field theory are equations for functionals like wave-function,
action functional, entropy and so on.

The equations of material systems are partial differential
equations for desired functions like a velocity of particles
(elements), temperature, pressure and density, which correspond to
physical quantities of material systems (continuous media). It had
been shown that from such equations it is obtained the
evolutionary relation for functionals (and state-functions) like
wave-function, action functional, entropy and others, in other
words, for functional of field theories. And this points to the
fact that the field-theory equations must be connected with the
evolutionary relation derived from the equations for material
systems.

If the nonidentical evolutionary relation be regarded as the equation 
for deriving 
identical relation with include closed forms (describing physical structures
desired), one can see that there is a correspondence between such evolutionary
relation and the equations for functional of existing field theories.
It can be verified that the equations of existing field theories
are either such equation or is analogous (differential or tensor)
to such equation. The solutions of field-theory equations are
identical relation obtained from nonidentical evolutionary
relation.

\bigskip

The results obtained show that when building the general field
theory it is necessary to take into account the connection of
existing field theories (which are based on the conservation laws
for physical fields) with the equations of noncommutative
conservation laws for material media (the balance conservation
laws for energy, linear momentum, angular momentum and mass
and the analog to such laws for the time, which takes into account
a noncommutativity of time and energy of material
system).

The theories of exterior and evolutionary skew-symmetric
differential forms, which reflect the properties of conservation
laws for physical fields and material media, allow to disclose and
justify the general principles of field theories and may serve as an 
approach to general field theory.

\bigskip

\rightline{\large\bf Appendix}

Below we present the table where 
physical fields and interactions in their dependence on the parameters $p$, $k$, $n$ of 
evolutionary and closed exterior forms are demonstrated. (Here $p$ is the degree
of evolutionary form in nonidentical relation, which is connected
with the number of interacting balance conservation laws,
$k$ is the degree of closed form generated by nonidentical
relation and $n$ is the dimension of original inertial space.)

This table corresponds to elementary particles.

{\footnotesize [It should be emphasized the following. Here the concept
of ``interaction" is used in a twofold meaning: the interaction of
balance conservation laws that relates to material systems,
and the physical concept of ``interaction" that relates to physical
fields and reflects interactions of physical
structures, namely, it is connected with exact conservation laws]}.

\bigskip
\centerline{TABLE}

{\scriptsize
\noindent
\begin{tabular}{@{~}c@{~}c@{~}c@{~}c@{~}c@{~}c@{~}}
\bf interaction&$k\backslash p,n$&\bf 0&\bf 1&\bf 2&\bf 3

\\
\hline
\hline
\bf gravitation&\bf 3&&&&
    \begin{tabular}{c}
    \bf graviton\\
    $\Uparrow$\\
    electron\\
    proton\\
    neutron\\
    photon
    \end{tabular}

\\
\hline
    \begin{tabular}{l}
    \bf electro-\\
    \bf magnetic
    \end{tabular}
&\bf 2&&&
    \begin{tabular}{c}
        \bf photon2\\
    $\Uparrow$\\
    electron\\
    proton\\
    neutrino
    \end{tabular}
&\bf photon3

\\
\hline
\bf weak&\bf 1&&
    \begin{tabular}{c}
    \bf neutrino1\\
    $\Uparrow$\\
    electron\\
    quanta
    \end{tabular}
&\bf neutrino2&\bf neutrino3

\\
\hline
\bf strong&\bf 0&
    \begin{tabular}{c}
    \bf quanta0\\
    $\Uparrow$\\
    quarks?
    \end{tabular}
&
    \begin{tabular}{c}
    \bf quanta1\\
    \\

    \end{tabular}
&
\bf quanta2&\bf quanta3

    \\
\hline
\hline
    \begin{tabular}{c}
    \bf particles\\
    material\\
    nucleons?
    \end{tabular}
&
    \begin{tabular}{c}
    exact\\
    forms
    \end{tabular}
&\bf electron&\bf proton&\bf neutron&\bf deuteron?
\\
\hline
N&&1&2&3&4\\
&&time&time+&time+&time+\\
&&&1 coord.&2 coord.&3 coord.\\
\end{tabular}
}

In the Table the names of the particles created are given. Numbers
placed near particle names correspond to the space dimension. Under the
names of particles the
sources of interactions are presented. In the next to the last row we
present particles with mass (the elements of material system) formed by
interactions (the exact forms of zero degree obtained by sequential
integrating the evolutionary relations with evolutionary forms of
degree $p$ corresponding to these particles). In the bottom row the
dimension of the {\it metric} structure created is presented.

From the Table one can see the correspondence between the degree $k$ of
closed forms realized and the type of interactions. Thus, $k=0$
corresponds to strong interaction, $k=1$ corresponds to weak interaction,
$k=2$ corresponds to electromagnetic interaction, and $k=3$ corresponds
to gravitational interaction.

The degree $k$ of closed forms realized and the number $p$ connected with the
number of interacting balance
conservation laws determine the type of interactions and the type
of particles created. The properties of particles are governed by the
space dimension. The last property is connected with the fact that
closed forms of equal degrees $k$, but obtained from the evolutionary
relations acting in spaces of different dimensions $n$, are distinctive
because they are defined on pseudostructures of different dimensions
(the dimension of pseudostructure $(n+1-k)$ depends on the dimension
of initial space $n$). For this reason the realized physical structures
with closed forms of degrees $k$ are distinctive in their
properties.

1. Petrova L.~I. Skew-symmetric differential forms: Conservation laws. 
Foundations of field theories. -Moscow, URSS, 2006, 158 p. (in Russian).  

2. Tolman R.~C., Relativity, Thermodynamics, and Cosmology. Clarendon Press,
Oxford,  UK, 1969.

3. Fock V.~A., Theory of space, time, and gravitation. -Moscow, 
Tech.~Theor.~Lit., 1955 (in Russian).

4. Clark J.~F., Machesney ~M., The Dynamics of Real Gases. Butterworths,
London, 1964.

5. Wheeler J.~A., Neutrino, Gravitation and Geometry. Bologna, 1960.

6. Tonnelat M.-A., Les principles de la theorie electromagnetique
et la relativite. Masson, Paris, 1959.

\end{document}